\documentclass[12pt,a4paper]{article}
\usepackage{graphicx,amsmath,amsfonts}
\def \ba{\begin{eqnarray}}\def\ea{\end{eqnarray}}
\def\bc{\begin{center}}\def\ec{\end{center}}
\def\nn{\nonumber\\}

\title{\huge \bf Impact of vector meson polarization on its interaction with matter.}
\author{\bf S.R.Gevorkyan, A.V. Guskov}

\begin{document}
\maketitle
\bc Joint Institute for Nuclear Research, 141980 Dubna, Russia \ec
{\bf Abstract}
The production of unstable particles on different nuclei provides the possibility to determine the total cross section of the interaction of vector mesons $V=\rho,\omega,\varphi, K^{*0}(892),J/\psi$ etc. with nucleons. This interaction is defined by a set of amplitudes that correspond to the transverse (helicity $\lambda=\pm 1$) or longitudinal ($\lambda=0$) polarization of the vector meson. The total cross section for the interaction of transversely polarized vector mesons with nucleon  $\sigma_T=\sigma(V_T N)$ has been extracted from the coherent photoproduction of vector mesons off nuclei, while the   vector meson production in charge exchange reactions as $\pi^\pm(K^\pm)+A\to  V^0(K^{*0})+A'$ provides the unique opportunity to obtain the not yet measured total cross section for longitudinally polarized vector meson interacting with nucleon $\sigma_L=\sigma(V_L N)$.\\
We shortly discuss  the importance of the knowledge of $\sigma_L$ and possibility to extract its value from experiments on nuclear target.

\section{Introduction}
The  forward scattering amplitude of a vector meson  on a nucleon  averaged over the nucleon spin is determined  by two quantities: $\sigma'_T=\sigma _T(1-i \alpha _T)$  and
$\sigma'_L=\sigma _T(1-i\alpha _L)$, where $\sigma_{T(L)}$ is the total cross sections for the interaction of a transversely (longitudinally) polarized vector meson with the nucleon and $\alpha_{T(L)}= Re f_{T(L)}(0)/Im f_{T(L)}(0)$ is the ratio of the real to  imaginary part of the corresponding amplitudes at zero angle. \\
The  vector meson forward scattering  amplitude  off  spinless target   reads:
\ba
f(\vec k,\vec k')=f_0(0) +f_1(0) (\vec S\vec n)^2
\ea
with $\vec S$  the spin of the vector meson and  $\vec n=\vec k/k$  the unit vector in the direction  $\vec k$.
 According to the optical theorem imaginary parts of complex functions  $f_0(0),f_1(0)$   can be expressed in terms of the corresponding total cross sections $\sigma_T,\sigma_L$:
 \ba
 Im f_0(0)=\frac{k}{4\pi} \sigma_L, ~~~  Im f_1(0)=\frac{k}{4\pi} \left(\sigma_T-\sigma_L\right)
\ea

The dependence of  vector particle interaction on its  polarization has been known for many years in the case when the constituents of the particle are  in the D-wave state. A good example of such dependence  is the deuteron interaction with matter~\cite{glauber69,faldt80}. The D-wave component in the deuteron wave function leads to the different absorption in the matter for transversely and longitudinally polarized  deuterons~\cite{dubna08,juelich10}.\\
Spin dichroism (dependence of interaction on particle polarization) leads to the appearance of tensor polarization~\cite{bar12}. The intensity of unpolarized deuteron beam ($I^0_{+1}=I^0_{-1} =I^0_0=1/3$) after it passage the distance z in the target with density $\rho$ depends on the value of total cross sections of deuteron interaction with atoms of the target  $\sigma_{\pm 1},\sigma_0$
\ba
I_{\pm 1}(z)=I^0_{\pm 1}e^{-\sigma_{\pm 1}\rho z};  I_0(z)=I^0_0 e^{-\sigma_0\rho z}
\ea
 The tensor polarization of the deuteron beam is determined by the difference $\sigma_0-\sigma_{\pm 1}$:
 \ba
 p_{zz}(z)=\frac{ I_{+1}(z)+I_{-1}(z) -2I_0(z)}{  I_{+1}(z)+I_{-1}(z) +I_0(z)}\approx \frac{2}{3}(\sigma_0-\sigma_{\pm 1})\rho z
 \ea
Thus the difference of  tensor polarization from the zero indicates that interaction of deuterons with target atoms depends on the deuteron polarization.\\
For instance,  in the case of scattering of the deuteron beam with energy E=5GeV on carbon target~\cite{dubna08}  this difference  is about  5\%, which is a  result of the presence of the D-wave  in  the deuteron wave function~\cite{tarasov08}.

As to the  case of vector mesons interaction  the presence  of nonzero orbital momentum between quarks is strongly correlated with the meson polarization~\cite{gerland98,ivanov06}.
Unlike the deuteron where the  contribution of the D-wave in the deuteron wave function is small, the D-wave contribution in the wave function of  relativistic vector meson  is noticeably. Exactly the  dominance of the D-wave part in the virtual photon wave function  at high virtuality  $Q^2$  allows to restore the scaling in deep inelastic scattering~\cite{ivanov06}.\\
The  quarks distribution in transversely and longitudinally polarized vector  mesons are very different. For instance the authors of ~\cite{ioffe01}  using the generalized QCD sum rules  showed that the distribution of valence quarks  in the longitudinally and transversely polarized $\rho$ meson diverse strongly depending on meson polarization.
The similar   distinction   takes place  for the distribution of constituent quarks in the vector mesons polarized transversely and longitudinally ~\cite{forshaw10,forshaw12}.\\
The difference  in quark distributions in vector mesons depending on its polarization  should  lead  to impact of vector meson polarization on their interaction  with the matter~\cite{gev16,gev17}.\\
To estimate  the values of $\sigma_T,\sigma_L$ we consider the vector meson scattering on nucleon in color dipole model~\cite{gev88,nikol91}. In any QCD description of photon or vector meson interaction with the target the first step is the conversion of the initial particle into a quark-antiquark pair. In color dipole model this pair interacts with a target as a color dipole, whose cross section depends  on the transverse separation r in the $q\bar q$ pair. The total cross section of the transverse and longitudinal polarized vector mesons with a target in mixed representation ( z-share of  the  meson  light-cone  momentum carried by the quark) reads:
\ba
\sigma^{L(T)}(VN)&=&\int |\Psi_V^{L(T)}(r,z)|^2\sigma(r)d^2rdz\nn
\sigma (r)&=&\sigma_0(s) \left (1-e^{-\frac{r^2}{R^2(s)}}\right)
\ea
Here $\Psi_V^{L(T)}(r,z)$ is the wave function (quark distribution) in the vector meson polarized transversally or longitudinally, whereas the simple parametrization of dipole cross section $\sigma(r)$ in (5) allows  to describe not only mesons photoproduction, but also the interaction of high energy hadrons with nucleons and nuclei~\cite{boris00}.\\
The dependence of $\sigma_L(\rho N)$ and $\sigma_T(\rho N)$ on invariant energy $W=\sqrt{s}$  calculated  by above expression using  parameterizations of the vector meson wave function in Boosted Gaussian approach~\cite{boris02} and ADS/QCD  Holographic model~\cite{forshaw10}  is shown in Fig.1. The impact of the polarization on the  total cross section of  the vector meson interaction with nucleon is large leading  to  the different absorption of vector mesons produced on nuclei depending on their  polarization~\cite{gev16,gev17}.\\
\begin{figure}[h]
\begin{center}
\includegraphics[width=0.6\linewidth,angle=0]{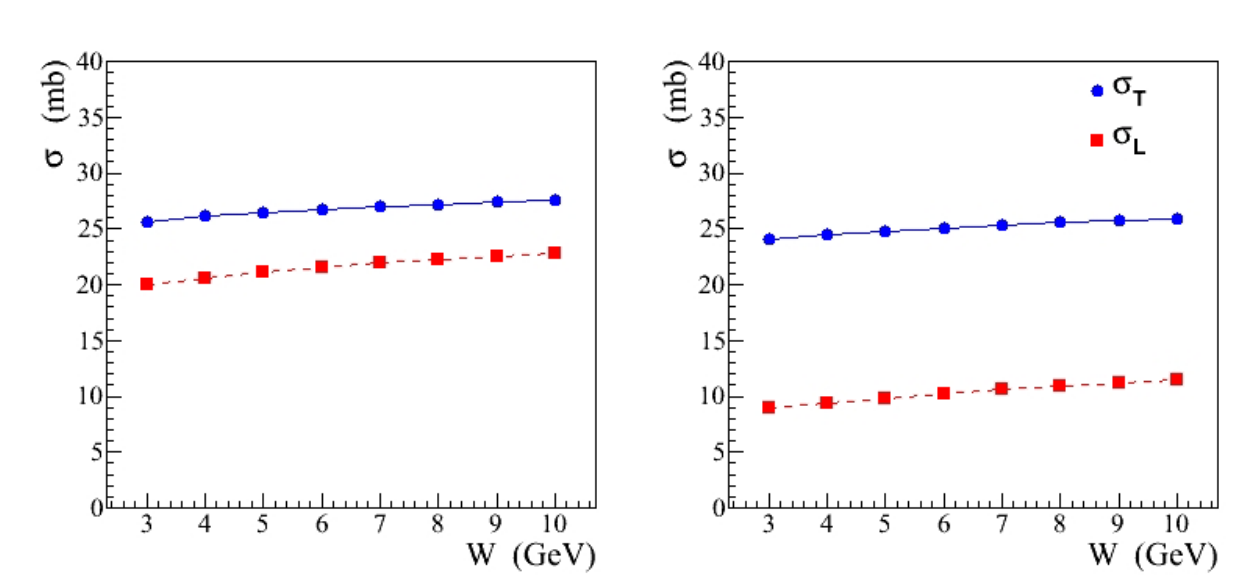}
\end{center}
\caption{  Longitudinal $\sigma_L$ and transverse $\sigma_T$  $\rho N$ total cross sections as a function of the invariant energy for different parameterizations    of the $\rho$  meson wave function:
(left)  Boosted Gaussian model~\cite{boris02} ; (right)   ADS/QCD  Holographic model~\cite{forshaw10}.}
\end{figure}
In view of vector mesons beams absence the best way to measure $\sigma_L(VN)$ and $\sigma_T(VN)$ are the measurement of vector mesons production on nuclei. For many years such processes are the source of unique information on interaction of unstable particles with nucleons~\cite{yennie78}.\\
In  vector mesons photoproduction due to the s-channel helicity conservation  $\rho,\phi$ mesons are produced mainly transversely polarized and only a part of $\omega$  mesons at JLAB energies (due to the pion exchange)  is longitudinally polarized~\cite{LOI15,chudakov16,proposal17}. Thus in experiments on vector mesons production by real photons off nuclei one extracts the total cross section of interaction of transversely polarized vector mesons with nucleons. To get information on interaction of longitudinally polarized vector mesons with nucleon one should measure the vector mesons production in processes where the longitudinally polarized vector mesons are mainly produced. Such kind of processes are vector mesons production on nuclei by pion or kaon beams $\pi^\pm(K^\pm)+A\to V^0(K^{*0})+A'$, where due to dominating pion exchange the produced vector mesons are mainly longitudinally polarized~\cite{brom84,AMBER18}. Later on we consider the vector mesons production off nuclei by pion and kaon beams with the aim to get the information on the value of $\sigma_L(VN)$ from such type of reaction. Before discussing this issue let us explain why the knowledge of $\sigma_L(VN)$ is important and in some cases even crucial.
\section{Vector meson polarization vs color transparency.}
The knowledge of the cross section  $\sigma_L(VN)$ is an important task for instance in interpreting the color screening effect in  the vector meson leptoproduction~\cite{gev17}.
The idea of the color transparency (CT) is that a hadron produced in certain hard-scattering processes has a smaller probability to interact in the nuclear matter due to its smaller size compared to the   physical hadron~\cite{fassi22}.
The vivid example of CT is the well known effect~\cite{yennie78} of "shrinking photon" in leptoproduction of vector mesons. At high photon virtuality  $Q^2$  the transverse  size of its hadronic component  $r\sim 1/Q^2$ is smaller than the size of a normal hadron. This would account for the pointlike behavior and the diminished absorption of virtual photons in nuclear matter. The  increase  with $Q^2$  the nuclear transparency defined as $Tr = \frac{\sigma_A}{A\; \sigma_N}$, where $\sigma_A$ and  $\sigma_N$ are vector mesons  production cross sections off  nuclei and nucleon respectively  considered as a direct result of CT effect~\cite{adams95,clas12}.\\
On the other hand   the increase of nuclear transparency with $Q^2$  can be partially  a result of the large difference (see Fig.1) between  $\sigma_L(VN)$ and $\sigma_T(VN)$.  Really the fraction of longitudinally
polarized vector mesons in leptoproduction  rises with  $Q^2$~\cite{her09}. Accounting that $\sigma_L(\rho)$ is much less than
$\sigma_T(\rho)$ the absorption of longitudinally polarized $\rho$ meson in nuclear matter should be weaker than for transverse one.\\
To estimate this effect we use the expression for nuclear transparency accounting for absorption of  vector mesons with different polarization~\cite{chudakov16}
\ba
Tr &=&\rho_{00}N(0,\sigma_L)+(1-\rho_{00})N(0,\sigma_T)\nn
N(0,\sigma)&=&\int\frac{1-\exp(-\sigma\int{\rho(b,z)dz})}{\sigma}d^2b,\nn
\ea
where  $\rho_{00}$ is the  spin  density matrix element corresponding to fraction of longitudinal polarized vector mesons production. In Fig.2 we cited the experimental data for nuclear transparency~\cite{clas12} and our calculations
denoted by stars accounting for different absorption of $\rho$ meson in nuclei depending on its polarization.In calculations we put $\sigma_T(\rho N)=25mb, ~~\sigma_L(\rho N)=10mb$. For the dependence of $\rho_{00}$ on $Q^2$  we used the  relation $\rho_{00}=\frac{\epsilon R}{(1+\epsilon R)}$ and the fit~\cite{her09} for ratio of longitudinal to transverse cross section on the nucleon $R(Q^2)=c_0\left(\frac{Q^2}{m^2}\right)^{c_1}$  with constants $\epsilon=0.8, c_0=0.56, c_1=0.47$ .\\
As seen from the Fig.2 the considered effect has to be separated from  the effect of color transparency as it is not small and exhibit the same behavior. Thus  the challenge of the impact of vector meson polarization on it interaction with nuclei and nucleon can be crucial for determination of such fundamental effect as color transparency.
\begin{figure}[ht]
\centering
\includegraphics[width=0.8\linewidth,angle=0]{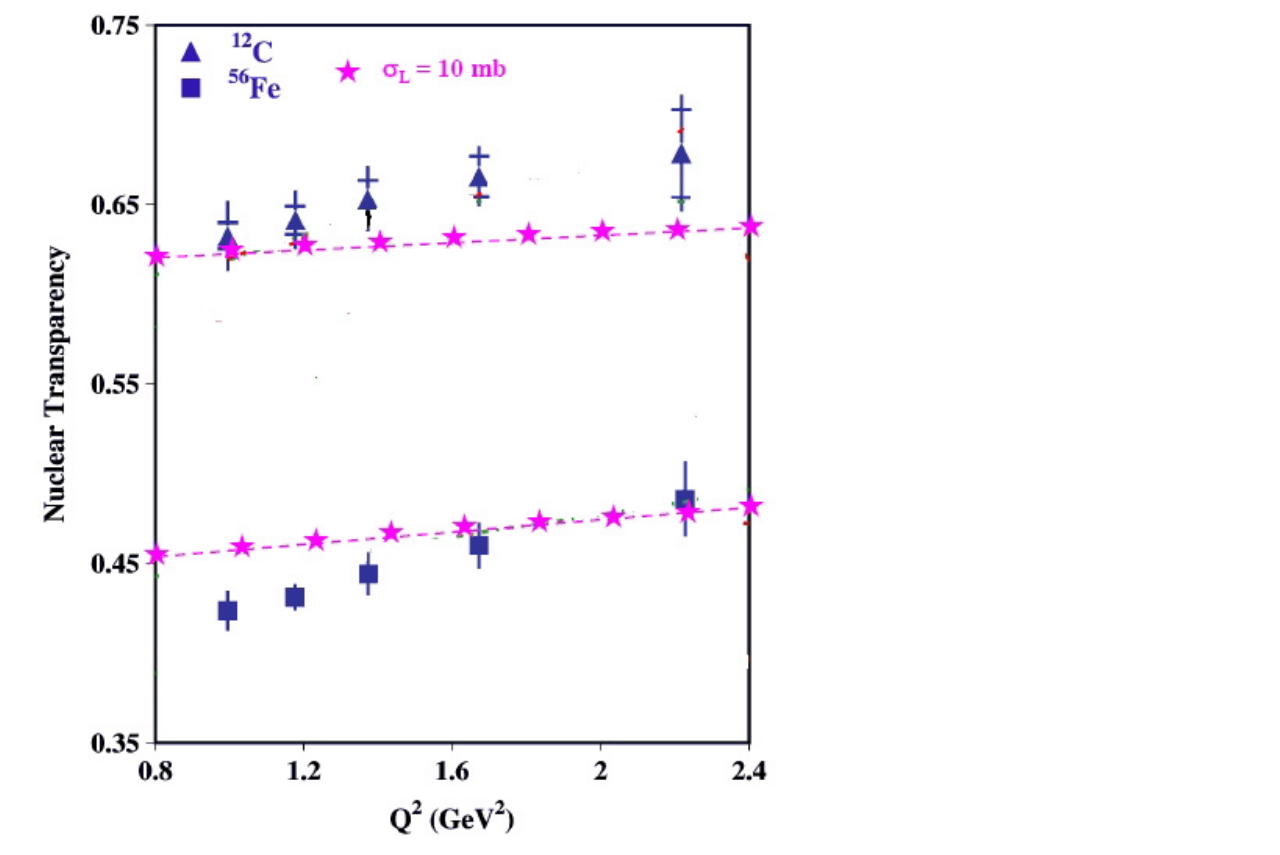}
\caption{ Nuclear transparency of $\rho^0$  electroproduction  as a function of  $Q^2$ compared to the experimental data  from CLAS~\cite{clas12}.}
\end{figure}

\section{Vector mesons production by pions.}
For the best of our knowledge,  the only attempt  to study the impact of the vector meson polarization on its  absorption in nuclei  was made many years ago ~\cite{itep78} using the charge exchange reaction  $\pi^- + A\to\rho^0+A^\prime$. The incoherent cross section and spin density matrix elements  of $\rho$  mesons were measured for  different nuclei: C, Al, Cu, Pb.

 Due to the dominance of the  pion exchange in this process, a large fraction of longitudinally polarized $\rho$ mesons was produced. At the first glance,  the experimental data support  the assumption that  $\sigma_T(\rho N)\approx\sigma_L(\rho N)$. However, there are strong reasons against  such a conclusion. It was  shown~\cite{tarasov75} that, due to the relatively low energy of the primary  pion beam ($E_{\pi} = 3.7\;{\rm  GeV}$) and  large decay width of the $\rho$ meson, the significant  part of  mesons decaying  inside the nucleus  complicates the  interpretation of the experimental data.

The total cross sections of the $\rho$ and $f(1270)$ mesons with a nucleon were measured at Argonne~\cite{argon73} using the  charge exchange process on neon nuclei
 $\pi^++Ne\to \rho (f)+Ne'$.   Accounting for  the possibility  of the $\rho$ mesons decay in nuclei ($p_{\pi} = 3.5$~GeV/$c$), the $\rho N$  total cross section is required to
be $\sigma(\rho N)\approx 12$ mb, which would  contradicts to the value   $\sigma(\rho N)\approx 27$ mb obtained from the $\rho$ meson photoproduction on nuclei.
From our point of view this difference is a direct  result of distinction between $\sigma_T(\rho N)$ and $\sigma_L(\rho N)$ as in charge exchange process  mainly  longitudinally
polarized $\rho$ mesons are produced,  whereas in the  photoproduction  $\rho$ mesons  are transversely polarized due to s-channel helicity conservation.

Taking into account that the vector meson decay mean free path in the laboratory system  grows with energy  $l=\frac{p}{m_V\Gamma_V}$,  the vector  mesons produced  by pion and kaon beams with  energies of tens GeV available at the M2 beam line at CERN~\cite{AMBER18},  would help  to determine uniquely the value of longitudinal $V N$  total cross section  $\sigma_L(V N)$ while  the value of the total cross section for transversely polarized vector mesons $\sigma_T(V N)$  is known from  photoproduction on nuclei~\cite{yennie78} and can be cross-checked at SPS energies.\\
Vector mesons production on protons by pions beams has been measured at high energies with the  goal to check the  Regge model predictions. It was shown that in $\omega$  and $\varphi$ mesons production ~\cite{om77,phi96} the main contribution at high energies give the natural parity exchange  leading to production of transversely polarized mesons. As  to the $\rho$ mesons production the dominance of pion exchange at modest transfer momenta leads to production predominantly the longitudinally polarized $\rho$ mesons~\cite{brom84}. Unlike  the pion exchange leads to the cross section falling  as  $1/p^2$, at energies relevant to AMBER experiment region the cross section of the reaction $\pi^-+A\to \rho+ A'$ is measurable and can give the unique  information on the value of
$\sigma_L(\rho N)$.\\
There are some plans to produce the RF-separated kaon beam for future measurements at the  experimental area of SPS~\cite{AMBER18,amber19}. If so  it is interesting to measure the charge exchange process $K^{\pm}+A\to K^*(892)+A'$  to get information on $K^*$  absorption in nuclear matter  depending on vector meson polarization.\\
 Recently ALICE collaboration presented the  data~\cite{alice20}  on  $K^*$  and $\varphi$ mesons production  in peripheral nucleus-nucleus collisions,  which show that vector mesons  polarization unlike their production in pp collisions  depends on their transverse momenta and centrality and can't be explained by conventional effects. Thus the investigation of $K^*$  production off nuclei becomes  a topical issue.
\section{Acknowledgements}
This work is dedicated to the memory of I.A. Savin who always supported and promoted our work and with whom we have a respect to work and associate for many years.

\end{document}